# Formation of $Cu_6Sn_5$ phase by cold homogenization in nanocrystalline Cu–Sn bilayers at room temperature

H. Zaka [a], S.S. Shenouda [a,b], S.S. Fouad [a], M. Medhat [c], G.L. Katona [b], A. Csik [d], G.A. Langer [b], D.L. Beke [b]

[a] Ain Shams University, Department of Physics, Faculty of Education, Roxy, Cairo, Egypt
[b] University of Debrecen, Department of Solid State Physics, Faculty of Science and Technology, H-4010, Debrecen, P.O. Box 2, Hungary
[c] Ain Shams University, Department of Physics, Faculty of Science, Cairo, Egypt
[d] Institute Nuclear Research, Hungarian Academy of Sciences (ATOMKI), H-4001, Debrecen, P.O. Box 51, Hungary

### abstract

Solid state reaction between nanocrystalline Cu and Sn films was investigated at room temperature by depth profiling with secondary neutral mass spectrometry and by X-ray diffraction. A rapid diffusion intermixing was observed leading to the formation of homogeneous $Cu_6Sn_5$ layer. There is no indication of the appearance of the $Cu_3Sn$ phase. This offers a way for solid phase soldering at low temperatures, i.e. to produce homogeneous $Cu_6Sn_5$ intermediate layer of several tens of nanometers during reasonable time (in the order of hours or less). From the detailed analysis of the growth of the planar reaction layer, formed at the initial interface in Sn(100 nm)/Cu(50 nm) system, the value of the parabolic growth rate coefficient at room temperature is $2.3 \times 10^{-15}$ cm$^2$/s. In addition, the overall increase of the composition near to the substrate inside the Cu film was interpreted by grain boundary diffusion induced solid state reaction: the new phase formed along the grain boundaries and grew perpendicular to the boundary planes. From the initial slope of the composition versus time function, the interface velocity during this reaction was estimated to be about 0.5 nm/h.

## 1. Introduction

Copper–tin (Cu–Sn) layers are of great technological and scientific interest, and are frequently used to avoid health and environmental problems caused by the use of lead-based alloys [1]. Sn thin films deposited on Cu-based substrate are often used for soldering of microelectronics devices [2]. According to the Cu–Sn phase diagram, there are two intermetallic phases which exist below 100 °C, $Cu_3Sn$ ($\epsilon$-phase) and $Cu_6Sn_5$ ($\acute{\eta}$-phase). The latter is richer in Sn content as compared to the former and is the only phase formed in bimetallic Cu–Sn films aged at room temperature [3–7].

The formation and growth of intermetallic compounds of Cu–Sn have been extensively studied [4–13]. In the pioneering works of Tu and his coworkers [3,14,15] the most important conclusions, obtained in thin film systems with Sn and Cu thicknesses in the range of 180–2500 nm, were as follows: i) The reaction between the Cu and Sn started spontaneously during the deposition at room temperature and led to the formation of $Cu_6Sn_5$ phase. Prolonged aging resulted in increase of amount of the (ordered) $Cu_6Sn_5$ phase. No $Cu_3Sn$ was detected even after very long annealing times. ii) Parallel with the formation of the $Cu_6Sn_5$ phase it compressed the Sn and led to the formation of whiskers on the free surface of Sn. iii) The $Cu_6Sn_5$ phase grew linearly with time, but it was mentioned [15] that "The linear growth … may suggest that the compound is not of layer type when the grains are very thin or small. The morphology of the $Cu_6Sn_5$ phase in the early stage of formation deserves more detailed studies." iv) From Rutherford backscattering investigations, using W markers, it was concluded that the dominant diffusing species is Cu. The phase grew into the Sn side and in their model of phase growth (see Fig. 6a in [15]) they neglected the diffusion of Sn into Cu arguing that in the growth of the phase both the interstitial diffusion of Cu in Sn and contribution from the transport along grain boundaries are important (the growth rate was slower in the thicker film [14]). The order of magnitude estimation of effective interdiffusion coefficient, related to the phase growth, gave $10^{-17}$–$10^{-16}$ cm$^2$/s values [14,15].

In addition to the above results it was concluded, from the analysis of the data obtained between 160 and 200 °C using also inert markers [10,16,17], that a) the intermetallic growth was related to Sn diffusion into the nanocrystalline Cu, which is in contrast to the point iv) above and b) the growth rates were about 10 times larger for nano-grained thin films than in coarse-grained massive samples.

Two more recent investigations also deserve attention. In [18] it was found that, in electroplated Cu/Sn bilayers the growth of the $Cu_6Sn_5$ phase was parabolic, mentioning that the linear growth undergoes to parabolic one with increasing the thickness of the growing layer. Furthermore, the $Cu_6Sn_5$ layer was continuous, but not uniform: at some locations it was very thin while at other positions relatively thick protrusions were observed. From the temperature dependence of the parabolic growth constant of the $Cu_6Sn_5$ phase K = $2 \times 10^{-15}$ cm$^2$/s can be estimated at room temperature.

In [19] a very careful and detailed investigation of solid state reaction between several microns thick Sn and polycrystalline, but bulk, Cu substrate was carried out at room temperature. The following observations were made: I) The $Cu_6Sn_5$ phase was formed, predominantly at the intersections of the Sn grain boundaries with the Sn/Cu interface, on the Sn side and grew only into Sn. During continuous aging the $Cu_6Sn_5$ proceeded significantly into the Sn grain boundaries, accompanied with a growth perpendicular to the grain boundaries (GBs). Upon prolonged aging the fine-grained $Cu_6Sn_5$ phase appeared on the face of the Sn grains adjacent to the Sn/Cu interface. Heterogeneous nucleation of these $Cu_6Sn_5$ islands and their subsequent coalescences led to the formation of a closed planar layer. II) In the interpretation of the above observations the authors of [19] expressed their opinion that the formation of the $Cu_6Sn_5$ phase at the Sn/Cu interface is the result of volume diffusion. After the formation of the compact $Cu_6Sn_5$ layer, the further planar growth of it was slower. At the same time the diffusion of Sn into the large grained Cu was neglected.

This study provides information concerning the effect of aging time on the growth kinetics and phase formation mechanism of $Cu_6Sn_5$ in the Cu–Sn nanocrystalline films. In contrast to previous studies, in our films both sides were really nanostructured and the application of the SNMS depth profiling is a novelty.

## 2. Experimental

Two sets of nanocrystalline Cu–Sn thin film samples of different layer thicknesses were prepared by DC magnetron sputtering in a vacuum of $10^{-7}$ mbar at room temperature. The films were obtained by consecutive sputtering of Cu followed by Sn onto (111) oriented SiN wafer that works as a barrier to prevent the Si diffusion in the system. Disk-shaped Cu and Sn targets with diameter of 2 in. were used as sputtering sources. The thicknesses of the layers were as follows: Sn(100 nm)/Cu(50 nm) and Sn(50 nm)/Cu(25 nm). The samples have the thickness ratio required for the formation of the intermetallic phase of $Cu_6Sn_5$ [3].

In the sputtering process a plasma discharge was maintained above the targets, sputtered onto the SiN substrate by Ar$^+$ ion bombardment (P = $5 \times 10^{-3}$ mbar). The deposition chamber was connected to ultra-high vacuum system (with $10^{-7}$ mbar). The sputtering chamber has three magnetrons with shutters. The nominal thicknesses of layers were estimated from the sputtering time and the predetermined deposition rate of each target. The sputtering rates were calculated from the layer thickness measured by the AMBIOUS XP-1 profilometer.

The Cu–Sn thin films were aged for various times. The concentration–depth profiles were measured by Secondary Neutral Mass Spectrometry, SNMS (INA-X, SPECS, GmbH, Berlin) [20,21], which works with noble gas plasma and the bombarding ion current has an extremely high lateral homogeneity. The low bombarding energies (in the order of 100 eV) and the homogeneous plasma result in an outstanding depth resolution (<2 nm) [22,23]. The SNMS data (intensity (cps)-time (s) spectra) were transformed into concentration–depth profiles, using the sensitivity factors of the elements and by using the proportionality between the intensity and the number of sputtered particles [24].

The structure of thin film couples and its changes due to diffusion and solid state reaction were studied by X-ray diffraction, XRD, using Siemens X-ray diffractometer, Cu-Kα X-rays of wavelength λ = 0.154 nm and the data were taken for the 2θ range between 20° and 60°.

## 3. Results

### 3.1. Phase formation in Sn(100 nm)/Cu(50 nm) films

Fig. 1a shows the intensity versus sputtering time of the as-deposited Sn(100 nm)/Cu (50 nm)/SiN sample. The intensity was converted to concentration using the sensitivity factors of the elements and the sputtering time was also converted to depth as shown in Fig. 1b. In the initial state, Cu is observed throughout the entire layer of Sn. Sn atoms are also observed through almost the whole Cu layer and only about a 15–20 nm thick Cu layer near the substrate is pure. This cannot be fully interpreted by the intensive GB diffusion during the deposition. The quite wide interface between Sn and Cu layers can also be attributed to the granulation of Sn as we observed from Scanning Electron Microscopic, (SEM) images, already on the surface of Sn (100 nm)/Cu (50 nm)/SiN (Fig. 2). Thus, Cu atoms could easily diffuse through the Sn layer and reached the Sn surface during the deposition.

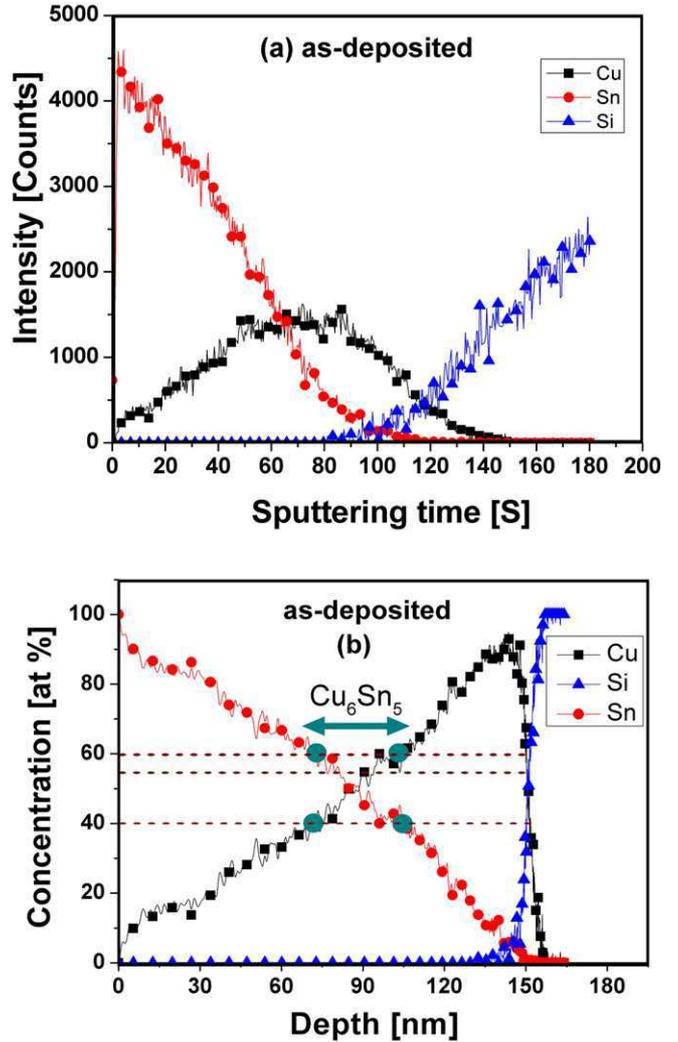

**Fig. 1.** SNMS profile of the as-deposited Sn(100 nm)/Cu(50 nm)/SiN sample; (a) Intensity versus sputtering time. (b) Concentration versus depth. The presence of intermetallic phase is indicated by a nearly flat region around the initial interface on the Cu and Sn curves (see also the text).

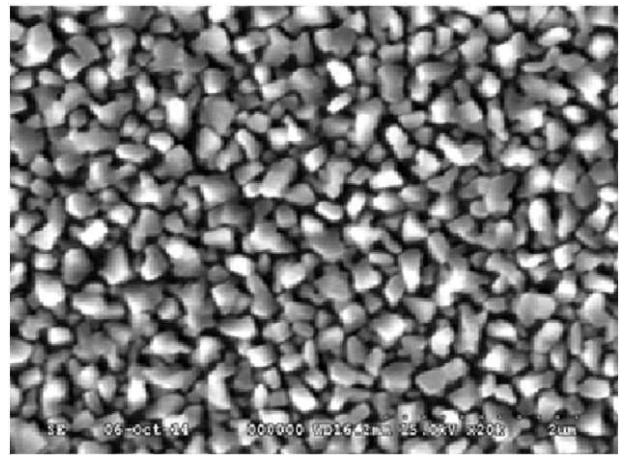

**Fig. 2.** SEM shows the top view of the as-deposited Sn(100 nm)/Cu(50 nm)/SiN sample. The average size of the granules is about 200 nm.

For the estimation of the region where the $Cu_6Sn_5$ phase is formed, one can assume a certain range of existence of this phase. The thickness of this concentration interval is suggested by the plateau region of about 20% wide in Fig. 1b. Thus on the SNMS profiles, we can draw a band of about ±20% wide. If we take the center of this line at about 55% of Cu (i.e. at the equilibrium composition of $Cu_6Sn_5$ phase), then the band, should be investigated, can be taken as 55% + 5%/−15%. It is a bit asymmetric, but it is often observed in diffusional solid state reactions that phases nucleate and grow with some deviations from the equilibrium compositions and in a wider existence range than those dictated by the phase diagram. From the crossing points between this interval and e.g. with the Cu curve, we can calculate the thickness of the intermetallic layer. According to Fig. 1b, SNMS data suggest the formation of the $Cu_6Sn_5$ phase as a layer with 33 nm thickness at the interface between the copper and tin films.

Fig. 3 shows the concentration–depth profiles at different aging times for the Sn(100 nm)/Cu (50 nm) film. Aging for 2 h (see Fig. 3a) leads to a slight redistribution of the components. The thickness of the $Cu_6Sn_5$ phase increases to about 56 nm. Increasing the aging time to 4 h, we observed the increase of the thickness of the intermetallic layer as shown in Fig. 3b and at the same time, the average composition also increases in the region outside the planar growth (see e.g. the Sn composition at 130 nm). It can also be seen that the average composition here further increases with increasing aging times and gradually approaches to the composition band of the $Cu_6Sn_5$ phase (Fig. 3c–e). Finally, at 533 h a homogeneous layer is obtained with average composition corresponding to $Cu_6Sn_5$.

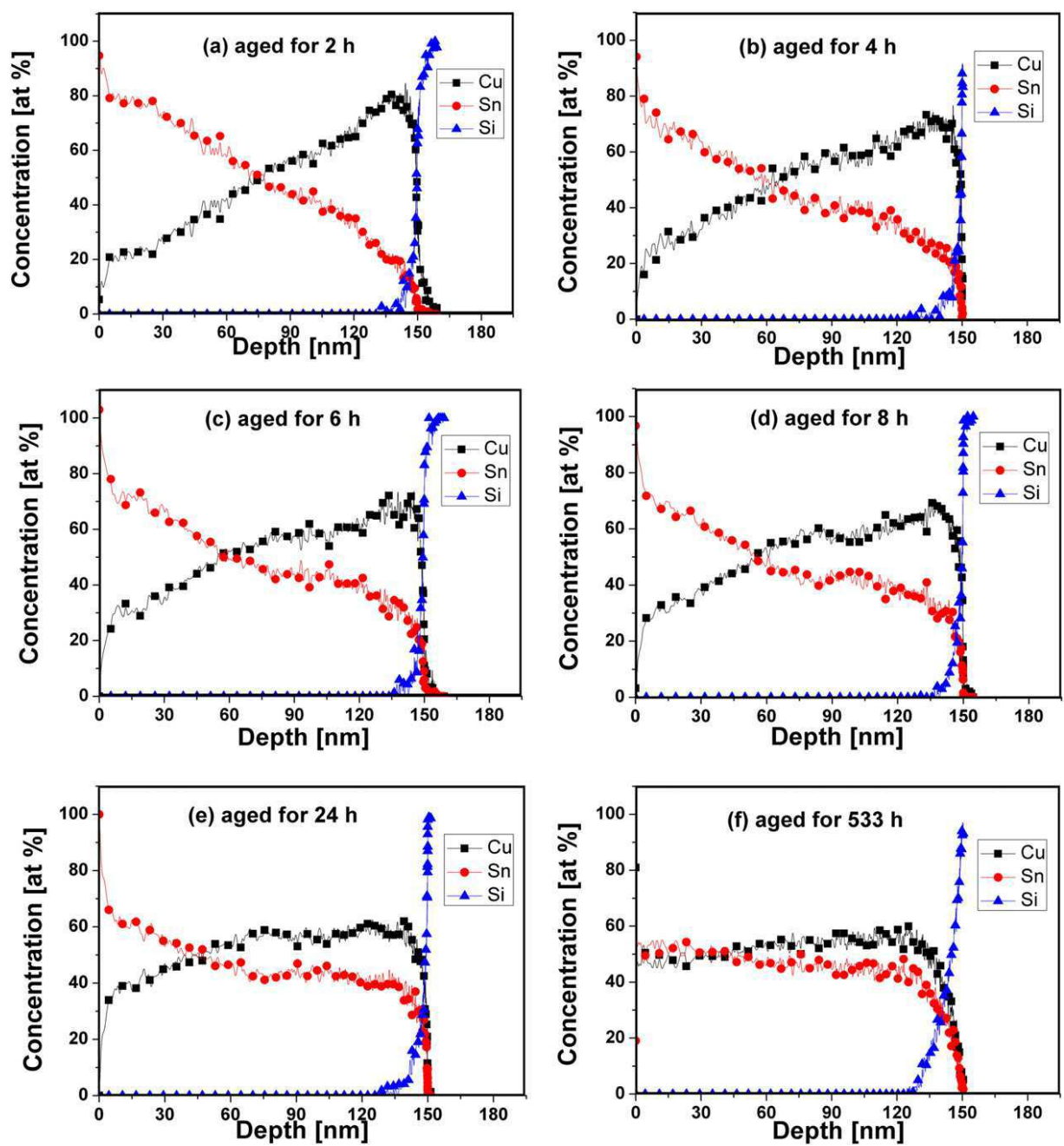

Fig. 3. Concentration versus depth in the Sn(100 nm)/Cu(50 nm)/SiN sample aged for 2, 4, 6, 8, 24 and 533 h.

The results of XRD measurements are shown in Fig. 4a and b. Fig. 4a shows the XRD pattern obtained on the as-deposited Sn(100 nm)/Cu(50 nm) film. This pattern shows peaks corresponding to Sn and Cu as expected. In addition, in accordance with the SNMS result (Fig. 1b) there are some peaks which correspond to $Cu_6Sn_5$, indicating the existence of a reaction between Cu and Sn at room temperature, either during sputtering or shortly afterwards.

Increasing the aging time to 533 h, a rapid growth of the ή phase was observed indicating an additional reaction between Cu and Sn (see Fig. 4b). Only diffraction peaks for $Cu_6Sn_5$ were detected which means that the whole film was transformed into the $Cu_6Sn_5$ layer.

Since the solubility of Cu in Sn and also Sn in Cu are negligibly small below 100 °C, no large lattice parameters change in Cu and Sn films is expected as intermixing takes place. Accordingly, no significant changes of the positions and shapes of Cu and Sn and $Cu_6Sn_5$ reflections were found. The grain sizes estimated from the full width at half maximum (FWHM) by the Scherrer equation are given in Table 1. Since the reflection of Cu was weak, we could not calculate its grain size.

**Table 1**
Results of grain size measurements for Sn (100 nm)/Cu (50 nm)/SiN sample, obtained by the use of Scherrer equation.

| Specimen | Aging time | Grain size |
| --- | --- | --- |
| Sn(100 nm)/Cu(50 nm)/SiN | As-deposited | Sn(39 nm) |
| | | $Cu_6Sn_5$(43 nm) |
| | After 533 h | $Cu_6Sn_5$ (30 nm) |

### 3.2. Phase formation in Sn(50 nm)/Cu(25 nm) films

Fig. 5 shows the SNMS depth profiles for Sn(50 nm)/Cu(25 nm) thin films at selected aging times. It can be seen that already in the as received state there is a considerable intermixing and even on the Cu side, near to the substrate, the Sn content is rather high. This can be the effect of granulation and the lower grain sizes present in the thinner films. Thus a detailed analysis of the growth processes is not possible. Nevertheless the final state is again the homogeneous $Cu_6Sn_5$ phase.

### 4. Discussion

It can be seen from the results presented above that, in accordance with [14,17], the nanocrystallinity has an effect on the kinetics of the solid state reaction (see e.g. Figs. 3e and 5c): faster process is observed in thinner films (with smaller grain size).

The second typical feature is the irregular growth morphology: parallel with the formation and growth of the compact reaction layer near to the initial interface a relatively fast increase of the Sn composition inside the Cu layer, far away from the initial interface, was detected. Qualitatively it is similar to the results described in [19]: there the formation of the $Cu_6Sn_5$ phase at the Sn grain boundaries, intersecting the Sn/Cu interface, was observed. These islands coalesced and formed a closed planar layer. Simultaneously "irregular" growth of the $Cu_6Sn_5$ phase along the Sn grain boundaries was observed.

Before giving a detailed discussion of our results, it is worth mentioning that there is an apparent contradiction between the results of [19] and ours, since they described the processes on the Sn side, while we treat the time evolution mainly on the Cu side. The reasons of this are as follows: i) in [19] the Cu was polycrystalline (but called "bulk" in order to express that the grain boundary diffusion could be neglected in Cu), while in our case both films were nanocrystalline (see also Table 1) and ii) in our case – partly also due to the complication related to the granulation of Sn grains during deposition – the diffusion was too fast on the Sn side and a detailed analysis is possible only on the Cu side. Note that our observations on the changes on the Cu side are in accordance with the results of [10] where nano-grained thin films were used.

Let us treat first the growth of the planar layer. As it was illustrated already in Fig. 1b, we calculated the time dependence of the thickness by the use of the composition band 55% + 5%/− 15%. Then using the relation,

$$X = (Kt)^n \tag{1}$$

one can calculate the growth exponent, n, and the growth rate coefficient K. Fig. 6a presents the X versus t relationship while Fig. 6b presents X versus $t^{1/2}$ dependence for the first part of the curve in Fig. 6a. It can be seen that – in accordance with the observation made on the as received sample (there exists a reaction layer already in this plot) – this curve has non zero intercept. From the value of the time given by the crossing point of this curve with the horizontal axis we can estimate the value of time correction: $t_{o1} = -1.44$ h. Thus for the determination of the exponent in Eq. (1) one has to use a corrected time scale $t' = t-t_{o1}$. In fact the origin of the time scale has to be shifted to the above crossing point, since the real time of the reaction is longer than the observation times in our experiment. Fig. 7a shows the log

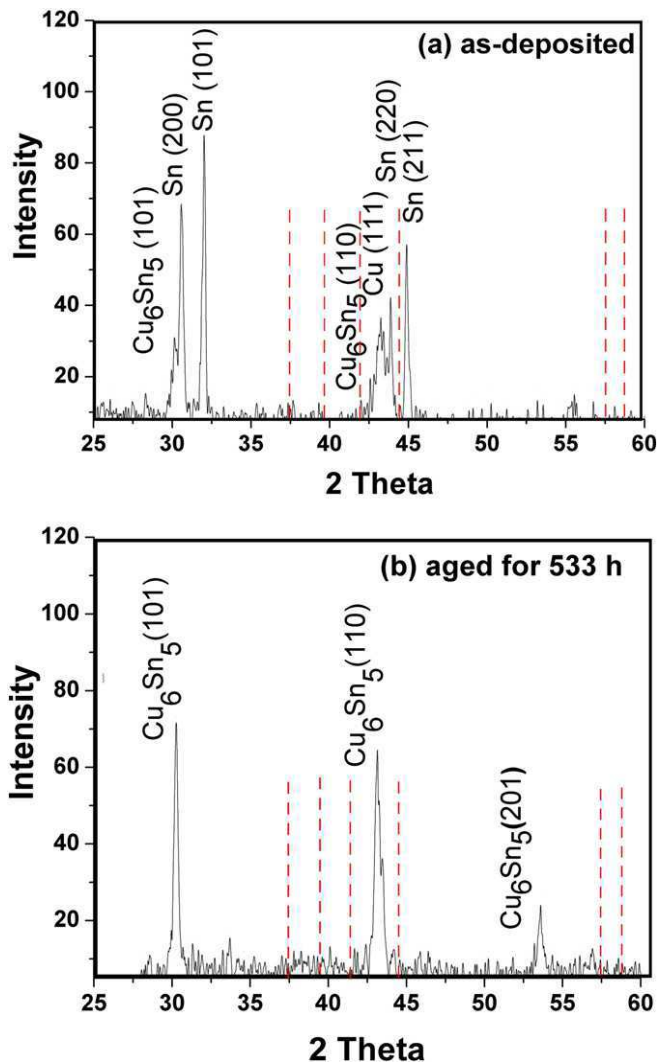

**Fig. 4.** X-ray diffraction for Sn(100 nm)/Cu(50 nm) films: (a) as-deposited (b) aged for 533 h. The dashed lines show the expected positions of $Cu_3Sn$: no $Cu_3Sn$ is detected.

89

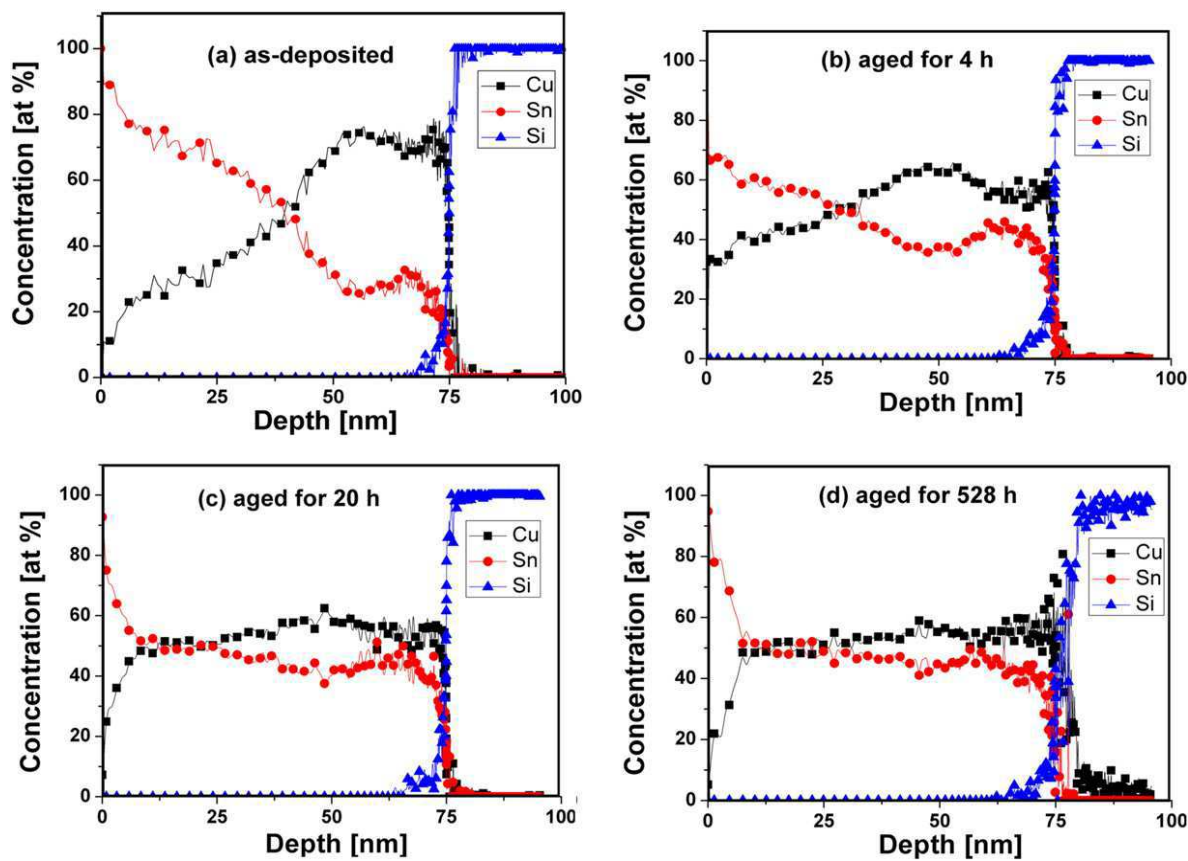

**Fig. 5.** Concentration versus depth of Sn(50 nm)/Cu (25 nm)/SiN sample for the as-deposited sample and the samples aged for 4, 20 and 528 h.

X versus log t′ function and the slope of the straight line fitted is n = 0.55, which is close to 0.5 expected for growth with diffusion control. In addition from the slope of the same function of X versus t′$^{1/2}$ plot (Fig.7b) the value of K can also be determined and it is K = 840 nm$^2$/h.

For the interpretation of the growth of the average composition near to the substrate in Cu, first we mention that in general diffusion along different grain boundaries can have important effects [12–28] on the overall intermixing process between two pure films. These processes can be well characterized by a bimodal GB network, with different (fast and slow) diffusivities. First the atoms migrate along fast GBs and accumulate at the film surface or at the film/substrate interface. These accumulated atoms form a secondary diffusion source for back diffusion along slow boundaries, which can lead to formation of a minimum on the average composition versus depth profiles close to the film/substrate interface (Fig. 5a and b). Thus the different GBs of the thin films can be gradually filled up with the diffusing atoms and the GB motion is perpendicular to the original boundary. The concentration–depth profiles reflect the result of these processes. Diffusion-induced grain boundary motion (DIGM) is an example of such kind of GB motions. Another process, which can also be related to the formation of a continuous reaction layer near to the initial interface, is the so called diffusion induced recrystallization, (DIR) which, besides the DIGM, is another manifestation of the relaxation of stresses accumulated by the inequality of the GB atomic fluxes. The planar growth of this layer can be the result of nucleation of the phase at the interface/GBs and a subsequent lateral growth of these nuclei by GB diffusion. If the product layer is not a solid solution but intermetallic compound then one can speak about GB diffusion-induced solid-state reaction (GBDIREAC) [29].

Thus during interdiffusion in binary systems where intermetallic layers can grow, it can be observed that the morphology of the formation and growth of the reaction product can be different from the usual picture observed at high temperatures, where the new continuous phases, formed at the initial interface, grew parallel to the contact surface only. Even the entire layer of the pure parent films can be consumed and a complete intermixing of components in the binary nanocrystalline couple can be achieved by grain boundary migrations through the volume [29]. Competition and/or simultaneous kinetics of the growth of the planar layer and the DIGM in the form of GBDIREAC can be also achieved. The above arguments can also explain the observed "irregular" growth described in [19].

It is worth mentioning that the X-ray diffraction data indicate that in the formation of the planar layer the role of DIR is probably less important as compared to nucleation and growth of nuclei at the intersection of the GBs. During DIR the new grains are usually much smaller than the initial grains and we came to the above conclusion because the grain size of the $Cu_6Sn_5$ is similar to the grain size of the Sn and it did not increase.

According to the above arguments we can summarize the following schematic picture for our processes, shown in Fig. 8, in accordance with the formation of the compact reaction layer at the original interface as well as with the growth of the average composition far from this interface.

In the region out of the planar growth (see e.g. the time evolution of the composition profiles at 130 nm depth) the increase of the average composition can be attributed to the GBDIREAC mechanism. Thus the $Cu_6Sn_5$ phase was formed along the grain boundaries of Cu and grew by the motion of the new interfaces. From the first stage of this time dependence of the average composition in this region, using the linear dependence of the average composition of Sn inside the Cu layer, the velocity of the grain boundary diffusion induced interface motion can also be estimated. Fig. 9 shows the average composition of Sn inside the Cu layer as a function of the aging time at 130 nm depth. Applying the equation, derived in [30,31], for the time evolution of the average concentration, $c_{Sn}$, in the nanocrystalline Cu layer,

$$c_{Sn}/c_e \cong 3\delta/d + 6vt/d, \qquad (2)$$

where $c_e$ is the equilibrium composition of the growing phase, which is $Cu_6Sn_5$ in our case, i.e. $c_e = 0.45$. d is the diameter of the grains in the Cu layer, $\delta$ ($= 0.5$ nm) is the GB thickness and v is the interface velocity. Eq. (2) can be obtained from a simple model of nanocrystal with spherical grains of d diameter with zero Sn inner concentration at the beginning and having a shell of $\delta/2$ thick around it. It is assumed that the shell has the composition of the equilibrium phase $Cu_6Sn_5$, $c_e = 0.45$ initially (say at $t_o$, necessary to fill in the GBs) and gradually widens due to the motion of the above shell, consuming the central part of the grain with a constant velocity, v [31]. For short times one arrives at linear expression (2) where $c_{Sn}$ is the actual composition obtained from the SNMS.

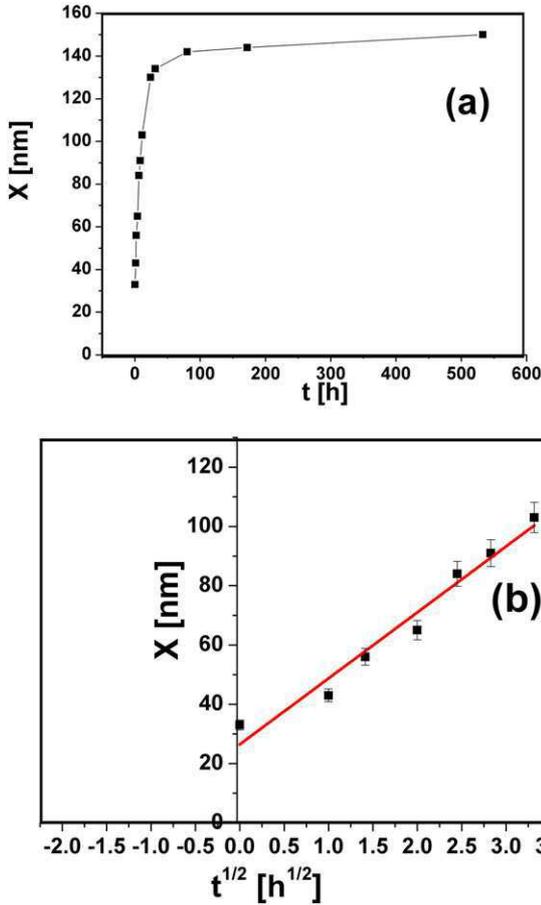

**Fig. 6.** a) Relation between the thickness of the intermetallic phase (X) and time of aging (t) for Sn(100 nm)/Cu(50 nm) film. b) X versus square root of time for the first 7 points shown in a).

For the use of Eq. (2) first one has to take into account that from the points shown in Fig. 9 only the first (linear) part has to be used and a new time correction, another than the one used for the application of Eq. (1), is necessary: $t'' = t - t_{o2}$. This takes into account that the motion of the above interface started at 130 nm depth before the SNMS depth profiling was made. The correction should fix the value of the intercept in Fig. 10, which shows the first, linear part of the plot shown in Fig. 9: it has to be $3\delta/d$ according to Eq. (2). The value of $3\delta/d$ can be estimated assuming that the grain size of the Cu is similar to the thickness of the film and to the grain size of Sn determined from the XRD data (Table 1): $d \cong 39$ nm and $\delta = 0.5$ nm, i.e. $3\delta/d = 0.04$. Thus the time scale has to be shifted accordingly. It is clear that $t_{o2}$ should be a relatively large value since the relative composition at 130 nm in the as received state is $c_{Sn}/c_e = 0.39$, larger than the intercept according to $3\delta/d = 0.04$. This requires $t_{o2} \cong -4.6$ h. From the slope of the $c_{Sn}/c_e$ versus the corrected time $t''$ function ($\cong 6v/d$), shown in Fig. 10, $v = 0.5$ nm/h is obtained.

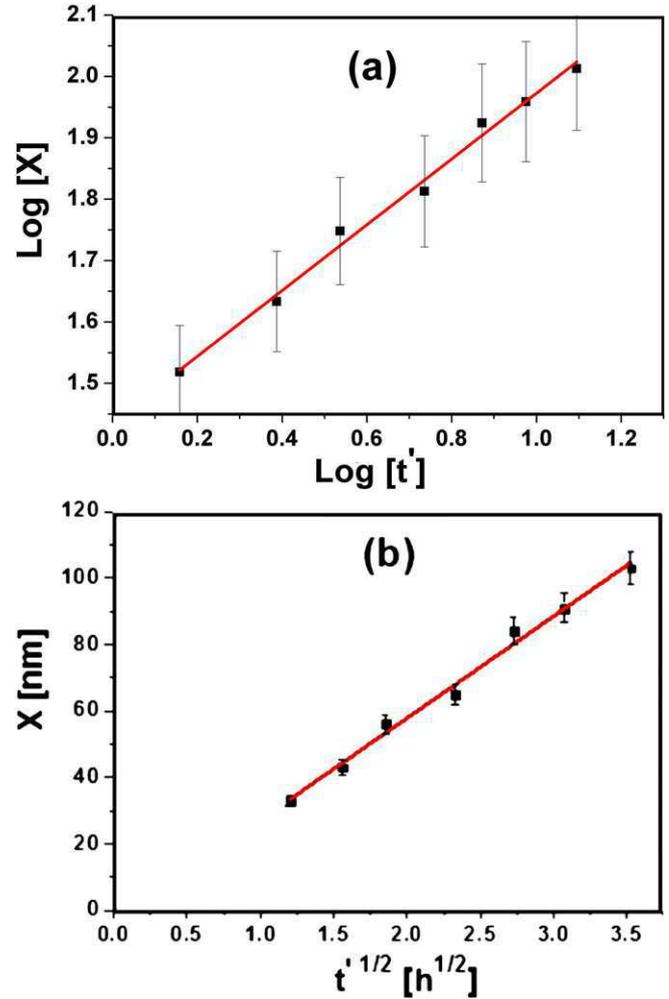

**Fig. 7.** a) Relation between Log (X) and Log (t′) for the Sn(100 nm)/Cu(50 nm) film with the corrected time scale, b) X versus square root of time with the corrected time scale.

The obtained value of the growth coefficient $K = 840$ nm$^2$/h $= 2.3 \times 10^{-15}$ cm$^2$/s should be compared with the effective diffusivities obtained in [14,15] ($10^{-17}$–$10^{-16}$ cm$^2$/s) and in [18] ($2 \times 10^{-15}$ cm$^2$/s), since our value as well as the values mentioned in the literature data should correspond to the effective diffusivity along the GBs of the $Cu_6Sn_5$ phase. It can be seen that our value is in a very good agreement with the value obtained in [18].

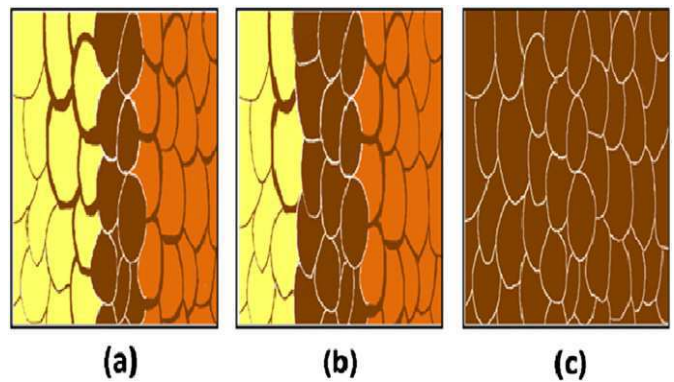

**Fig. 8.** Schematic illustration of the time evolution of DIGM and the planar growth. A and B are yellow and orange, respectively and the reaction layer is in brown. (a) As-deposited state, (b) intermediate aging time and (c) final state.

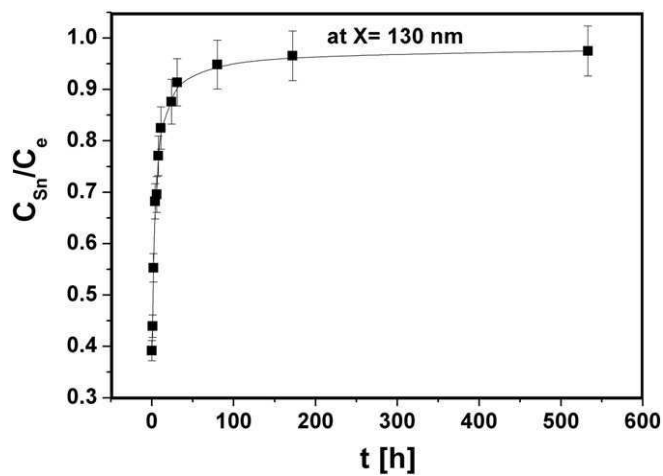

Fig. 9. Relation between the average composition of Sn inside the Cu layer and the aging time at depth 130 nm.

Furthermore, the interface velocity determined by us (0.5 nm/h) can be compared to the value obtained in [14] for the linear growth constant, k, in the Sn(200 nm)/Cu(560 nm) sample: k = 6 nm/h. It can be seen that our value is about one order of magnitude less than the k obtained from the overall increase of the amount of the $Cu_6Sn_5$ phase in the Sn. Thus our results as expected indicate that the velocity of the interface between $Cu_6Sn_5$ and Cu is slower than that of between Sn and $Cu_6Sn_5$ (i.e. these processes are slower on the Cu side). It is worth mentioning that in the beginning of the process linear kinetics is expected (with constant interface velocity) and at longer times the parabolic growth of the compact layer formed at the original interface becomes more rate controlling, especially if one measures the time dependence of the overall amount of the growing phase.

Finally, we have to mention that the formation of the compact layer at the Sn/Cu interface is too fast to be interpreted by bulk diffusion as it was proposed in [19]. Instead we suggest – similarly as it was explained e.g. in [30] for solid state reaction between $Ni_2Si$ and Si – that the formation of nuclei is caused by a DIR process and the simultaneous grow of these grains (and nucleation of further new grains) by GB diffusion can explain the growth of the continuous reaction layer. This also explains why the parabolic growth constant is related to the atomic transport along GBs.

## 5. Conclusions

The general features of the obtained results are as follows:

- In nanocrystalline Sn/Cu bilayers of different thicknesses the solid state reactions at room temperature can result in formation of homogeneous $Cu_6Sn_5$ phase, recommended for soldering.
- There is no indication on the appearance of the $Cu_3Sn$ phase.
- The shorter times necessary for the above phase formation in nanocrystalline bilayers offers a way for solid phase soldering at low temperatures, i.e. to produce homogeneous $Cu_6Sn_5$ intermediate layer of several tens of nanometers during reasonable time (in the order of hours or less).
- From the detailed analysis of the growth of the planar reaction layer formed at the initial interface in Sn(100 nm)/Cu(50 nm) films it was obtained that the value of the growth rate coefficient at room temperature is K = 840 nm$^2$/h = $2.3 \times 10^{-15}$ cm$^2$/s.
- The overall increase of the composition near to the substrate in Cu of Sn(100 nm)/Cu(50 nm) films was interpreted by grain boundary diffusion induced solid state reaction, GBDIREAC, and from the initial slope of the composition versus time function the interface velocity during this reaction was estimated to be about 0.5 nm/h.

### Acknowledgments

This work was done according to the agreement between the Faculty of Education, Ain Shams University (Cairo–Egypt) (Coordinator Prof. Dr. Suzan Fouad) and the Faculty of Science and Technology, Debrecen University (Debrecen–Hungary) (Coordinator Prof.Dr. Dezső Beke). The project was funded by the (OTKA) Grant No. NF CK80126 as well as by the European Union and the state of Hungary, co-financed by the European Social Fund in the framework of the TÁMOP 4.2.4. A/2-11-1-2012-0001'National Excellence Program' (author G.L. Katona) and TÁMOP- 4.2.2.A-11/1/KOV-2012-0036 projects.

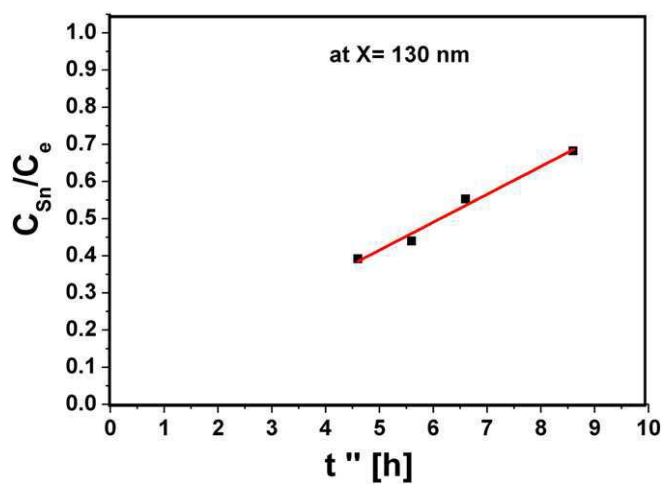

Fig. 10. Linear relation between the average composition of Sn inside Cu at depth 130 nm and the aging time with the corrected time scale for the first four points shown in Fig. 9.